\documentclass[aps, preprint]{revtex4-2}
\usepackage{amsfonts}
\usepackage{amsmath}
\usepackage{amssymb}
\usepackage{graphicx}
\usepackage{braket}
\usepackage{bbm}
\usepackage{microtype} 

\usepackage{tikz}
\usetikzlibrary{calc,positioning,patterns}

\usepackage[justification=Justified]{caption}
\usepackage{xcolor}
\usepackage[normalem]{ulem} 

\begin{document}
\title{Quantum simulacra}
\author{L. F. Alves da Silva and M. H. Y. Moussa.}
\affiliation{Instituto de Física de São Carlos, Universidade de São Paulo, P.O. Box 369, São Carlos, 13560-970, SP, Brazil}

\begin{abstract}
Here we analyze the creation of quantum simulacra: phenomena that emerge from treating a Hermitian or non-Hermitian quantum system in metrics other than the standard $L^{2}$. Changing the metric redefines the set of system observables and thus the experimental arrangement for their measurement, making quantum contextuality and microscopic reality metric-dependent. The simulacra therefore consist, on the one hand, of a resizing of the status of quantum measurement, which has always occupied a central role in quantum mechanics: beyond the connection between quantum and classical dynamics, measurements performed in an appropriate metric can emulate a microscopic reality distinct from that prescribed by the Hamiltonian. On the other hand, simulacra provide a route to implementing quantum operations that lie beyond the reach of the $L^2$ metric.  Quantum simulacra offer, as an example, an explanation for the recent observation of the violation of Bell inequalities with unentangled photons [Sci. Adv. \textbf{11}, eadr1794 (2025)]: photons that are separable in $L^2$ metric, become entangled when analyzed within a new metric framework. Simulacrum comes at the cost of implementing measurements of the metric-redefined observables; to address this challenge, we propose a scheme combining positive operator-valued measures with postselected subensembles.
\end{abstract}

\maketitle
\newpage

\section{Introduction}

Around the turn of the 21st century, the physics community saw the emergence of pseudo-Hermitian quantum mechanics, which is now a prolific area of research. In 1998, Bender and Boettcher \cite{BB} showed that the family of non-Hermitian Hamiltonians $H=p^{2}-(ix)^{N}$ can have a purely real, positive spectrum when parity-time ($\mathcal{PT}$) symmetry is unbroken ($N\geq 2$). Under spontaneous $\mathcal{PT}$-symmetry breaking---i.e., when $H$ and the $\mathcal{PT}$ operator no longer share a common eigenbasis, despite $\left[H,\mathcal{PT}\right]=0$---two real eigenvalues coalesce at an exceptional point and then emerge as a complex-conjugate pair \cite{Bender99}. Soon after, Mostafazadeh \cite{M} presented a framework that, in addition to the real spectrum, ensures norm conservation for pseudo-Hermitian dynamics. The key ingredient is a positive-definite metric operator that defines a modified Hilbert-space inner product, distinct from the standard $L^{2}$, with respect to which the non-Hermitian Hamiltonian becomes self-adjoint.

Since then, many theoretical and experimental developments have been reported in this field. Exceptional points associated with the $\mathcal{PT}$ phase transition have been exploited to uncover a wide range of phenomena \cite{PTPT}, and Mostafazadeh’s framework for autonomous non-Hermitian systems has been extended to nonautonomous settings in Refs.~\cite{Znojil,GW,FM}. While in autonomous systems $\mathcal{PT}$-symmetry breaking is witnessed by the eigenenergy spectrum \cite{BB}, in nonautonomous systems it manifests through the dynamical and geometrical phases: the symmetry is observed when the real and imaginary parts of these phases are odd and even functions of time, respectively \cite{TDSBreaking}. Time-dependent antilinear symmetries for time-dependent pseudo-Hermitian Hamiltonians have also been investigated, along with a symmetry-metric connection \cite{BeyondPT}.

Returning to the foundations of quantum mechanics, the Copenhagen interpretation connects, through the collapse of the wave function, the act of measurement to the emergence of quantum reality. More recently, this interpretation has been formalized in the concept of quantum contextuality \cite{QC}, which claims that quantum phenomena depend on the specification of the experimental apparatus and which observables are measured together. Unlike in classical theories, it is not possible to assign pre-existing values to the properties of the system, independent of the context of the experimental setup. In quantum simulacra, the redefinition of the metric results in redefined observables and thus the \textit{quantum contextuality}. The appropriate choice of metric can then shape the desired reality through the act of measurement. Simulacra then resize the role of quantum measurement, with microscopic reality showing itself to be plastic, unlike the rigidity of classical reality.

A recent experiment reporting a violation of Bell inequalities with unentangled photons \cite{WangZeilinger} provides a particularly straightforward example of this metric dependence in experimentally accessible phenomena. As emphasized in Ref. \cite{WW}, the reported Bell violation originates from two elements: a postselected subensemble and an unconventional normalization procedure. As we show below, these elements are precisely the ingredients required to define an nonstandard metric. More specifically, postselection determines the sector in which the measurement of observables is performed, while normalization fixes the inner product associated with that metric, with respect to which the correlations are evaluated. The result is therefore not a violation of Bell inequalities by unentangled states in the standard $L^{2}$ metric, which is not observed in a standard Bell test, but a simulacrum of Bell violation: a set of quantum correlations generated by evaluating a postselected subensemble with respect to a metric other than $L^2$.

Beyond its conceptual implications, the possibility of reshaping microscopic reality through metric redefinition opens a route to the implementation of quantum operations that lie beyond the scope of the standard $L^2$ metric. We stress the possibility of creating from the Hamiltonian-metric connection, unusual nonlinear amplifications or interaction processes that are absent from the $L^2$-Hermitian Hamiltonian. While Hamiltonians, together with the experimental setup,  define quantum contextuality in standard quantum mechanics, here we introduce metrics as an additional ingredient for shaping reality.

As a first step towards simulacra, in Section II, we adapt Mostafazadeh’s framework, originally formulated for non-Hermitian Hamiltonians, to Hermitian ones. This step circumvents the technically challenging task of engineering non-Hermitian Hamiltonians. In Section III, we present a method for measuring pseudo-Hermitian observables based on a combination of positive operator-valued measure (POVM) and postselection. Whereas the choice of metric is mandatory for the treatment of non-Hermitian Hamiltonians, for Hermitian ones it becomes an additional instrument for constructing microscopic reality. Starting from Section IV, we illustrate this idea with examples showing that phenomena absent from the Hamiltonian can emerge from the choice of metric, including effective interactions that enable entanglement generation and quantum-state teleportation in otherwise noninteracting settings. In particular, in Section VI, we show that simulacra fully account for the recent observation of Bell inequality violations with unentangled photons \cite{WangZeilinger}. Our conclusions are presented in Section X.

\section{Metrics for Hermitian and non-Hermitian Hamiltonians}

In this section, we consider an extension of the concept of pseudo-Hermiticity to Hermitian Hamiltonians, which requires redefining the metric and, consequently, observables of the Hermitian system in the analogous manner as for non-Hermitian systems. Starting from the Schrödinger equation for a non-Hermitian $H$, $i\partial_{t}\ket{\psi}=H\ket{\psi}$, and transforming it through the Dyson map $\eta$, we obtain $i\partial_{t}\ket{\phi}=h\ket{\phi}$, where $h=\eta H\eta^{-1}$ is the isospectral partner of $H$ and $\ket{\phi}=\eta\ket{\psi}$. By requiring $h$ to be Hermitian with respect to the standard $L^2$ metric, we obtain the pseudo-Hermiticity condition
\begin{equation}
\Theta H=H^{\dagger}\Theta , \label{PHc}
\end{equation}
which demands a nonunitary map $\eta$ and a Hermitian metric operator $\Theta=\eta^{\dag}\eta$. Pseudo-Hermiticity also implies pseudo-unitarity of the time-evolution operator, $U=\exp(-iHt)$, namely $U^{\dagger}\Theta=\Theta U^{-1}$ \cite{pseudoU}. Moreover, we verify that the eigenvalue equation $h\ket{\phi_{n}}=\varepsilon_{n}\ket{\phi_{n}}$ unfolds into
\begin{equation}
H\ket{\psi_{n}}=\varepsilon_{n}\ket{\psi_{n}} \quad \text{and} \quad  H^{\dag}\ket{\chi_{n}}=\varepsilon_{n} \ket{\chi_{n}}, \label{BBe}
\end{equation}
where the eigenstates $\ket{\psi_{n}}=\eta^{-1} \ket{\phi_n}$ and $\ket{\chi_{n}}=\eta^\dagger \ket{\phi_n}$ define a biorthogonal basis for the pseudo-Hermitian $H$, satisfying
$\braket{\chi_m|\psi_n}=\delta_{mn}$ and
$\sum_{n}\ket{\psi_{n}}\bra{\chi_{n}}=\mathbbm{1}$. The pseudo-Hermitian observables related to $H$ are given by $\mathcal{O}=\eta^{-1}o\eta$ with $o$ being usual Hermitian observables. Under the new
metric $\Theta$, the expectation value of the operator $\mathcal{O}$ is given by
\begin{equation}
\left\langle \psi(t)\left\vert \mathcal{O}\right\vert \psi(t)\right\rangle
_{\Theta}=\left\langle \psi(t)\left\vert \Theta\mathcal{O}\right\vert
\psi(t)\right\rangle =\left\langle \phi(t)\left\vert o\right\vert
\phi(t)\right\rangle,
\label{O-ev}
\end{equation}
with the unitarity of evolution assured by
$\left\langle \psi(t)\right\vert \left.  \psi(t)\right\rangle _{\Theta}
=\left\langle \phi(t)\right\vert \left.  \phi(t)\right\rangle $.

Now, applying the same reasoning to a Hermitian Hamiltonian, the pseudo-Hermiticity relation reduces to $\left[\Theta,H\right]=0$. Thus, $\Theta$ can be identified as an invariant and also as the generator of a continuous symmetry of the Hamiltonian. If, in the case of a non-Hermitian $H$, the two eigenvalue equations (\ref{BBe}) define the biorthogonal basis, for the Hermitian $H$ the equivalent equations,
$H\ket{\psi_{n}}=\varepsilon_{n}\ket{\psi_{n}}$ and
$H\ket{\chi_{n}}=\varepsilon_{n} \ket{\chi_{n}}$, define two sets of stationary states and, consequently, two generally distinct solutions of the Schrödinger equation due to the invariance of $\Theta$, namely $\ket{\psi(t)}$ and $\ket{\chi(t)}=\Theta\ket{\psi(t)}$. The remaining observables, besides the Hamiltonian $H$, are still given by $\mathcal{O}=\eta^{-1}o\eta$, and measurements of these observables, whose expectation values are defined by Eq.~(\ref{O-ev}), can lead to the emergence of simulacra.

\section{A protocol for measuring pseudo-Hermitian observables}

Measuring a pseudo-Hermitian observable $\mathcal{O}$ requires an experimental setup that distinguishes its eigenstates $\left\{ \ket{\tilde{\chi}_{n}} ,\ket{\tilde{\psi}_{n}} \right\} $. A measurement performed on the normalized solution of the Schrödinger equation $\ket{\psi}$ can be described by the biorthogonal projectors $\ket{\tilde{\psi}_{n}}\bra{\tilde{\chi}_{n}}$, which are measurement operators \cite{Brody}  that transform the quantum state $\ket{\psi}$ into a post-measurement state $\ket{\psi_{n}}$ with probability $p_{n}=\big|\braket{\tilde{\psi}_{n}\vert\psi}_{\Theta}\big|^{2}$. To implement this measurement, here we introduce a scheme that combines POVM with postselection, where the POVM operators are $E_{n}=q\ket{\tilde{\chi}_{n}}\bra{\tilde{\chi}_{n}}$ and $E_{\times}=\mathbbm{1}-\sum_{n}E_{n}$, where $q>0$ is chosen such that $E_\times\ge0$. POVMs provide a generalized measurement framework that extends standard projective measurements \cite{NC-qc}. This formalism is often required in quantum information theory, particularly for the optimal discrimination of non-orthogonal quantum states.

For our purposes, we consider a postselected subensemble defined by discarding all outcomes corresponding to $E_{\times}$. Otherwise, we would be performing the POVM measurement on the standard $L^2$ metric instead of $\Theta$. In fact, by performing the postselection, the outcome $E_n$ occurs $N_n$ times out of $N$ accepted measurements, so that: (i) the normalized probability is $p_{n}=\big| \braket{\tilde{\psi}_{n}\left|\Theta\right|\psi} \big|^{2}=\lim_{N\rightarrow\infty}N_{n}/N$ and, consequently, (ii) the expectation value of $\mathcal{O}$ is $\left\langle \psi|\mathcal{O}|\psi \right\rangle _{\Theta}=\sum_{n}o_{n}p_{n} $, as required in the $\Theta$-metric space. 

The proof of (i) follows from the standard POVM measurement postulate. Before postselection, the probability of obtaining the outcome associated with $E_{n}$ is $\left\langle \psi\left|E_{n}\right|\psi\right\rangle /\left\langle \psi\vert\psi\right\rangle =\lim_{N_{0}\rightarrow\infty}N_{n}/N_{0}$, where $N_{0}=N+N_{\times}$ is the total number of measurement runs. Postselection discards the $N_{\times}$ outcomes associated with $E_{\times}$, so that the probability in the accepted subensemble is $p_{n}=N_{n}/(N_{0}-N_{\times})=N_{n}/N$. Equivalently, since $N_{\times}=N_{0}\left\langle \psi\left|E_{\times}\right|\psi\right\rangle /\left\langle \psi\vert\psi\right\rangle $, the same conditional probability can be written as $p_{n}=\left\langle \psi\left|E_{n}\right|\psi\right\rangle /(\left\langle \psi\vert\psi\right\rangle -\left\langle \psi\left|E_{\times}\right|\psi\right\rangle )$. Using the completeness relation, which implies $\sum_{n}\left|\tilde{\chi}_{n}\right\rangle \left\langle \tilde{\chi}_{n}\right|=\Theta$, together with the normalization $\left\langle \psi\left|\Theta\right|\psi\right\rangle =1$, we have $E_{\times}=\mathbbm{1}-q\Theta$ and therefore $p_{n}=\big|\braket{\tilde{\psi}_{n}\vert\psi}_{\Theta}\big|^{2}$, which is the Born rule in the $\Theta$-metric space \cite{Brody}.  From this procedure, we obtain a postselected subensemble described by the metric $\Theta$, extracted from a larger ensemble (without postselection) under the standard $L^2$ metric.

\section{Squeezing mechanism}

To construct a simulacrum of the squeezing mechanism, we start from the Hamiltonian
\begin{equation}
H=\omega\left(a^{\dagger}a+1/2\right)  + \mu \left(\alpha a+\beta a^{\dagger}\right),
\end{equation}
which describes a radiation-field mode with frequency $\omega$ subject to a non-Hermitian linear drive. Here, $\mu\in\mathbb{R}$ sets the drive strength, while the real parameters $\alpha$ and $\beta$, with $\alpha\neq\beta$, characterize the non-Hermitian deformation. We introduce the Dyson map  $\eta=\exp\left[\epsilon a^{\dagger}a+\chi\left(\alpha a^{2}-\beta a^{\dagger2}\right)\right]$, which ensures the pseudo-Hermiticity condition (\ref{PHc}), once $\epsilon$ and $\chi$ are fixed in terms of a free parameter $z$, as
\begin{subequations}
\begin{align}
\epsilon(z) &  =\frac{1}{\sqrt{1+z^{2}}}\operatorname{atanh}\left[\frac{\alpha-\beta}{\alpha+\beta}\sqrt{1+z^{2}}\right], \\
\chi(z) &  =\frac{z\epsilon(z)}{2\sqrt{\alpha\beta}},
\end{align}
\end{subequations}
where each admissible value of $z$ in the interval $-z_{0}<z<z_{0}=2\sqrt{\alpha\beta}/\left(\alpha-\beta\right)$ then gives rise to a distinct metric.

The Dyson map $\eta$ is chosen within $\mathfrak{su}(1,1)$ algebra, rather than within the Heisenberg--Weyl algebra associated with the linear terms in $H$, so that the metric itself induces the parametric amplification underlying the squeezing mechanism. The resulting Hermitian counterpart of $H$ is given by 
\begin{equation} 
\begin{aligned} 
h={} & \omega\frac{1+\zeta^{2}}{1-\zeta^{2}}\left(a^{\dagger}a +1/2\right) +\mu\sqrt{\alpha\beta\frac{1+\zeta}{1-\zeta}} \left(a +a^{\dagger}\right) \\ &+\frac{\omega\zeta}{1-\zeta^{2}} \left(a^{\dagger 2}+a^{2}\right).
\end{aligned}
\end{equation}
where $\zeta=z/z_{0}\in\left(-1,+1\right)$. In contrast to $H$, the Hermitian $h$ describes a parametrically amplified field giving rise to the squeezing mechanism.

The pseudo-Hermitian quadrature observables are defined as $X_{\ell}=\eta^{-1}x_{\ell}\eta$ where $x_{1}=\left(a+a^{\dagger}\right)/\sqrt{2}$ and $x_{2}=\left(a-a^{\dagger}\right)/\sqrt{2}i$ are the standard Hermitian quadratures. For the normalized vacuum state, their uncertainties read $\Delta X_{1}=e^{-r}/\sqrt{2}$ and $\Delta X_{2}=e^{+r}/\sqrt{2}$, with the squeezing parameter $r=\tanh^{-1}\zeta$. The sign of $\zeta$ determines which quadrature is squeezed: for $\zeta>0$, $\Delta X_{1}<1/\sqrt{2}$, whereas for $\zeta<0$, squeezing occurs in $X_{2}$. Remarkably, $r$ diverges as $\zeta\rightarrow\pm1$, implying an unbounded degree of squeezing at the boundaries of the metric-parameter domain.

\subsection{Squeezing from a Hermitian free mode}

In the Hermitian linear-drive limit, $\alpha=\beta$, the above parametrization collapses to the trivial case, $\epsilon=\chi=0$ and $\zeta\rightarrow0$ for any finite $z$. Hence, the Dyson map defined above does not produce a quadratic squeezing term in $h$ in this limit. By contrast, if one sets $\mu=0$, the Hermitian counterpart $h$ still retains a quadratic contribution, and one of the quadratures is squeezed exactly as in the pseudo-Hermitian Hamiltonian setting described above. Thus, squeezing can arise even from a free radiation-field mode, through the choice of metric rather than through explicit quadratic terms in the original Hermitian Hamiltonian $H=\omega\left(a^{\dagger}a+1/2\right)$, a paradigmatic example of what we call a quantum simulacrum.

\subsection{Protocol for measuring the pseudo-Hermitian quadratures}

We now adapt our ``POVM + postselection" scheme introduced in Section II to observables with continuous spectra, such as the field quadratures. We define the continuous POVM operators as $E(x_{\ell})=qM^{\dagger}(x_{\ell})M(x_{\ell})$ and $E_{\times}=I-\int dx_{\ell}\,E(x_{\ell})$, with $\ell=1,2$ , where we take the measurement operator to be $M(x_{\ell})=\eta^{-1}\left|x_{\ell}\right\rangle \left\langle x_{\ell}\right|\eta$ with $\ket{x_{\ell}}$ denoting the eigenstates of Hermitian field quadratures. The probability density for obtaining the outcome $x_{\ell}$ is $p(x_{\ell})=\left\langle \psi\left|E(x_{\ell})\right|\psi\right\rangle /\braket{\psi|\psi}$. As in the discrete case discussed above, our scheme requires a postselection that can be implemented by a unitary gate $U_{\times}$, satisfying $U_{\times}E_{\times}U_{\times}^{\dagger}=\ket{0}\bra{0}$, so that the failure event is mapped onto the vacuum state. The resulting protocol, sketched in Fig.~1, supplements Leonhardt’s eight-port homodyne measurement scheme with a success-failure mechanism \cite{Leonhardt}. In the setup of Fig.~1(a), an event is discarded whenever the detection following the $U_\times$ gate yields the vacuum state; all other events are retained.The normalized probability density in the postselected subensemble is therefore $p(x_{\ell})=\left|\left\langle X_{\ell}\vert\psi\right\rangle _{\Theta}\right|^{2}$, with $\ket{X_{\ell}}=\eta^{-1}\ket{x_{\ell}}$, and the expectation value of the pseudo-Hermitian quadrature follows as $\left\langle \psi\left|X_{\ell}\right|\psi\right\rangle _{\Theta}=\intop dx_{\ell}\,x_{\ell}p(x_{\ell})$.

\begin{figure*}[t] 
\centering \includegraphics[width=\textwidth,height=100mm,keepaspectratio]{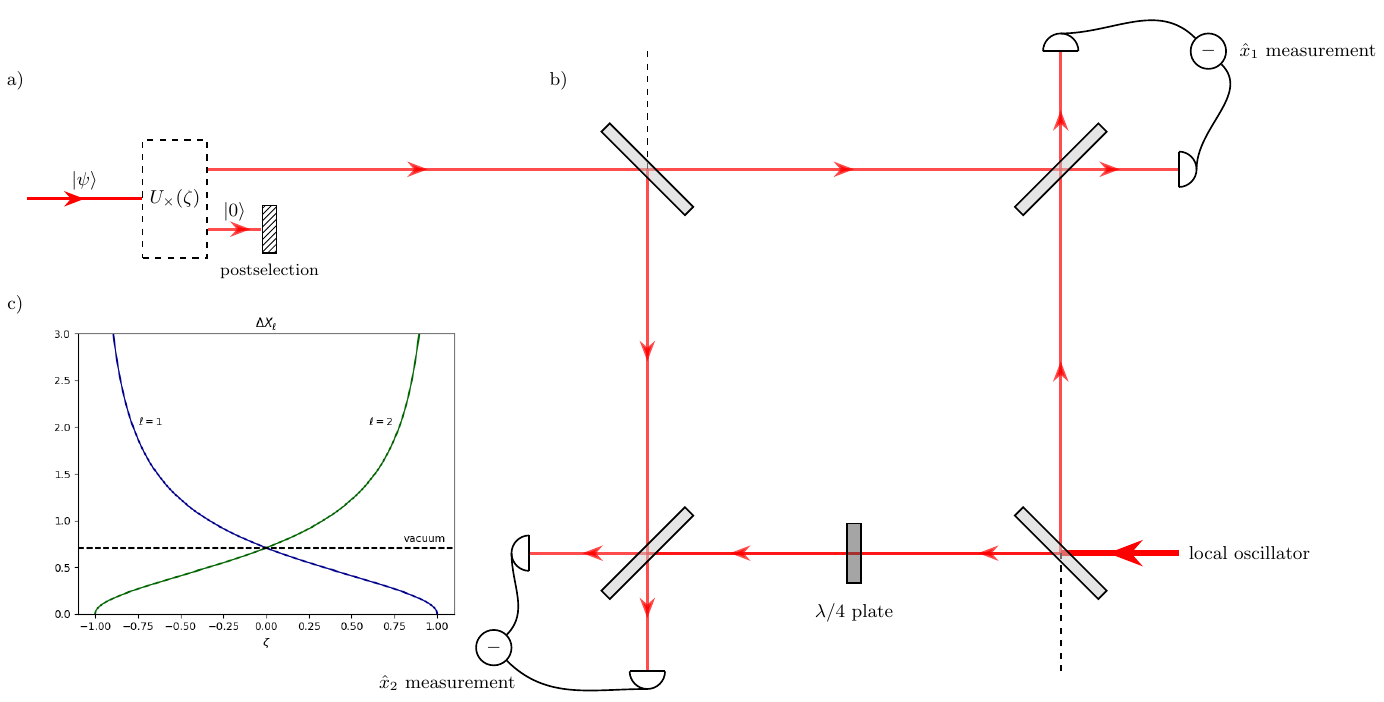} \caption{(a) Success-failure mechanism implementing the postselection required by the continuous “POVM + postselection” scheme. (b) The postselected state enters an eight-port homodyne interferometer allowing the simultaneous measurement of the quadrature $x_1$ and $x_2$ from the difference photocurrents. (c) The theoretical uncertainties of the pseudo-Hermitian quadratures shown as a function of $\zeta$.}
\end{figure*}

The postselected state enters an eight-port homodyne detector shown in Fig.~1(b), allowing a simultaneous measurement of the canonically conjugate quadratures $x_1$ and $x_2$. The balanced homodyne detector is the main component of this Leonhardt's apparatus, which uses a 50:50 beam splitter to interfere the signal mode $\left|\psi\right\rangle $ with a coherent state $\left|\gamma_{\text{LO}}\right\rangle $ of a strong local oscillator (i.e., $\ensuremath{\gamma_{\text{LO}}=\left|\gamma_{\text{LO}}\right|e^{i\varphi}}$ with $\left|\gamma_{\text{LO}}\right|\gg1$). The two outputs are then detected as photocurrent differences, $\Delta I$, resulting in a value proportional to the field quadrature $x_{\varphi}$ set by the local oscillator phase: $\Delta I\propto\left|\gamma_{\text{LO}}\right|x_{\varphi}$ with $x_{\varphi}=\left(a^{\dagger}e^{i\varphi}+a e^{-i\varphi}\right)/\sqrt{2}$. The eight-port detector contains two balanced homodyne detectors that operate in parallel, with both the signal and the local oscillator splitting into two arms. One of these arms is shifted by a relative phase of $\pi/2$ (implemented by a $\lambda/4$ plate), enabling simultaneous measurement of two non-commuting quadratures: $x_{1}\propto\Delta I_{1}/\left|\gamma_{\text{LO}}\right|$ and $x_{2}\propto\Delta I_{2}/\left|\gamma_{\text{LO}}\right|$ with $\varphi=0$. The theoretical uncertainties of the pseudo-Hermitian quadratures are shown in Fig. 1(c), where $\Delta X_{\ell}$ is plotted as a function of $\zeta$.

\section{Entanglement simulacrum}

To construct a simulacrum of bipartite entanglement, we start from the Hermitian Hamiltonian 
\begin{equation}
H=\omega_{1}\sigma_{1}^{z}+\omega_{2}\sigma_{2}^{z}, \label{H-ent}
\end{equation}
which describes two uncoupled two-level systems with transition frequencies $2\omega_{1}$ and $2\omega_{2}$, assuming a positive detuning $\Delta=\omega_2-\omega_1$ for simplicity. We introduce a Dyson map
\begin{equation}
\eta=\exp\left[\epsilon\sigma_{1}^{z}+\chi\left(\alpha\sigma_{1}^{+}\sigma_{2}^{-}-\beta\sigma_{1}^{-}\sigma_{2}^{+}\right)\right], \label{DMEnt}
\end{equation}
which carries the bilinear interaction that, although absent from the Hamiltonian, is required for the metric-induced construction of entanglement between the two subsystems. The pseudo-Hermiticity condition is satisfied provided that 
\begin{subequations}
\begin{align}
\epsilon(z) &  =- \frac{1}{\sqrt{1-z^{2}}}\operatorname{atanh}\left(\frac{\alpha-\beta}{\beta+\alpha}\sqrt{1-z^{2}}\right), \\
\chi(z) &  =\frac{z\epsilon(z)}{\sqrt{\alpha\beta}},
\end{align}
\end{subequations}
where $\alpha$, $\beta$, and $z$ are free parameters selecting the metric operator under the constraints $-1<z<1$ and $\alpha\beta>0$. The Hermitian counterpart of $H$ then reads
\begin{equation}
h=\frac{\omega_{1}+\omega_{2}\zeta^{2}}{1+\zeta^{2}}\sigma_{1}^{z}+\frac{\omega_{2}+\omega_{1}\zeta^{2}}{1+\zeta^{2}}\sigma_{2}^{z}-\frac{2\Delta\zeta}{1+\zeta^{2}}\left(\sigma_{1}^{+}\sigma_{2}^{-}+\sigma_{1}^{-}\sigma_{2}^{+}\right), \label{h-ent}
\end{equation}
where we have introduced the real parameter $\zeta=z\left(\alpha-\beta\right)/2\sqrt{\alpha\beta}$. Note that $\zeta=0$ yields vanishing effective coupling, whereas $\zeta=1$ corresponds to a maximal effective interaction. Starting from the separable superposition state
\begin{equation}
\left|\psi(0)\right\rangle =\left(a_{1}\left|\uparrow_{1}\right\rangle +b_{1}\left|\downarrow_{1}\right\rangle \right)\otimes\left(a_{2}\left|\uparrow_{2}\right\rangle +b_{2}\left|\downarrow_{2}\right\rangle \right), \label{EntIniSta}
\end{equation}
with the normalization condition $\langle \psi(0)|\Theta|\psi(0)\rangle=1$, namely
\begin{equation}
\left[a_{1}^{\ast}a_{1}e^{2\epsilon(z)}+\left(\alpha/\beta\right)b_{1}^{\ast}b_{1}\right]\left[a_{2}^{\ast}a_{2}e^{2\epsilon(z)}+\left(\beta/\alpha\right)b_{2}^{\ast}b_{2}\right]=e^{+2\epsilon(z)},
\end{equation}
we solve the Schrödinger equation for $h$ by expanding the corresponding initial state $\ket{\phi(0)}=\eta\ket{\psi(0)}$ in the eigenbasis of $h$. The corresponding state in the Hermitian representation evolves as
\begin{equation}
\begin{aligned}
\ket{\phi(t)}
& =
e^{\epsilon}a_1a_2 e^{-i\Omega t}
\ket{\uparrow_1\uparrow_2}
+
e^{-\epsilon}b_1b_2 e^{+i\Omega t}
\ket{\downarrow_1\downarrow_2}
\\
&+
\frac{\sqrt{{\beta}/{\alpha}}\,a_1b_2 e^{+i\Delta t}
-
\zeta\sqrt{{\alpha}/{\beta}}\,b_1a_2 e^{-i\Delta t}}{\sqrt{1+\zeta^2}}
\ket{\uparrow_1\downarrow_2}
\\
&+
\frac{\zeta\sqrt{{\beta}/{\alpha}}\,a_1b_2 e^{+i\Delta t}
+
\sqrt{{\alpha}/{\beta}}\,b_1a_2 e^{-i\Delta t}}{\sqrt{1+\zeta^2}}
\ket{\downarrow_1\uparrow_2}.
\end{aligned}
\label{phi-ent}
\end{equation}
This allows us to compute the expectation values of pseudo-Hermitian observables. For instance, the spin observable of subsystem 1 is $S_{1}^{z}=\eta^{-1}\left(s_{1}^{z}\otimes I_{2}\right)\eta$ and its expectation value evolves as 
\begin{equation}
\left\langle S_{1}^{z}(t)\right\rangle =\mathcal{A}+\mathcal{B}_{c}\cos\left(2\Delta t\right)+\mathcal{B}_{s}\sin\left(2\Delta t\right)
\end{equation}
where
\begin{subequations}
\begin{align}
\mathcal{A} &= \frac{1}{2} [ |a_{1}a_{2}|^{2}e^{2\epsilon} - |b_{1}b_{2}|^{2}e^{-2\epsilon} + \frac{ \left(\alpha^{2}|a_{2}b_{1}|^{2} -\beta^{2}|a_{1}b_{2}|^{2}\right) \left(\zeta^{2}-1\right) }{ \alpha\beta\left(1+\zeta^{2}\right) } ], \\
\mathcal{B}_{c} &= \frac{2\zeta}{1+\zeta^{2}} \operatorname{Re} \left[ \left(a_{1}b_{2}\right)^{\ast} \left(a_{2}b_{1}\right) \right], \\
\mathcal{B}_{s} &= \frac{2\zeta}{1+\zeta^{2}} \operatorname{Im} \left[ \left(a_{1}b_{2}\right)^{\ast} \left(a_{2}b_{1}\right) \right]. 
\end{align}
\end{subequations}
The oscillatory terms appear only if both initial atomic states contain local coherence, i.e., $a_i b_i\neq0$, which is converted by the effective coupling into an observable signature of the entangling dynamics. 

To quantify entanglement, we consider concurrence, which for a pure two-qubit state reads $\mathcal{C}=|\braket{\phi|\sigma_{y}^{(1)}\otimes\sigma_{y}^{(2)}|\phi^{\ast}}|=|\braket{\psi|\eta^{-1}\sigma_{y}^{(1)}\otimes\sigma_{y}^{(2)}\eta^{\ast}|\psi^{\ast}}_{\Theta}|$, where $\ket{\phi^{\ast}}$ denotes the complex conjugate of $\ket{\phi}$. The concurrence witnesses the emergence of two-qubit entanglement \cite{Wootters}, satisfying $0\le\mathcal{C}\le1$: $\mathcal{C}=0$ if and only if $|\phi\rangle$ is separable, whereas $\mathcal{C}=1$ for maximally entangled states. To demonstrate that separability depends on the choice of metric, we compute the concurrence of the state $\ket{\phi(t)}$, obtaining
\begin{equation}
\mathcal{C}(t)=\frac{2\left|\zeta\right|}{1+\zeta^{2}}\left|2\zeta a_{1}a_{2}b_{1}b_{2}-\left(\alpha/\beta\right)a_{2}^{2}b_{1}^{2}e^{2i\Delta t}+\left(\beta/\alpha\right)a_{1}^{2}b_{2}^{2}e^{-2i\Delta t}\right|,
\end{equation}
which is modulated by a pre-factor that allows for maximum entanglement $\mathcal{C}=1$ at $\mathcal{\zeta}=\pm1$, as for example with $a_1=b_2=0$.

Following the measurement scheme introduced above for pseudo-Hermitian observables under the metric $\Theta$, we introduce the POVM operators $E_{n}=q  \left|\chi_{n}\right\rangle \left\langle \chi_{n}\right|$ and $E_{\times}=I-\sum_{n}E_{n}$ where $\left|\chi_{n}\right\rangle $ denotes the biorthogonal eigenvectors of $S_{1}^{z}$, given by 
\begin{subequations}
\begin{align}
\ket{\chi_{\uparrow\uparrow}} & =e^{+\epsilon}\ket{\uparrow_1\uparrow_2}, \\
\ket{\chi_{\uparrow\downarrow}} & = \frac{\beta\left|\uparrow_{1}\downarrow_{2}\right\rangle -\alpha\zeta\left|\downarrow_{1}\uparrow_{2}\right\rangle }{\sqrt{\alpha\beta}\sqrt{1+\zeta^{2}}}, \\
\ket{\chi_{\downarrow\uparrow}} & = \frac{\alpha\left|\downarrow_{1}\uparrow_{2}\right\rangle +\beta\zeta\left|\uparrow_{1}\downarrow_{2}\right\rangle }{\sqrt{\alpha\beta}\sqrt{1+\zeta^{2}}}, \\
\ket{\chi_{\downarrow\downarrow}} & = e^{-\epsilon}\left|\downarrow_{1}\downarrow_{2}\right\rangle .
\end{align}
\end{subequations}
The experimental implementation of the ``POVM + postselection" scheme requires a global two-qubit operation capable of discriminating the biorthogonal basis associated with $S_1^z$. For instance, such a measurement can be constructed using an entangling gate such as a CNOT between spin 1 (control) and spin 2 (target), $\mathrm{CNOT}=\left\vert \uparrow_{1}\right\rangle \left\langle \uparrow
_{1}\right\vert \otimes I_{2}+\left\vert \downarrow_{1}\right\rangle
\left\langle \downarrow_{1}\right\vert \otimes\sigma_{2}^{x}$. In the conventional case, entanglement generation occurs through time evolution governed by a bilinear interaction between two qubits. In the simulacrum, this bilinear interaction is absent from the Hamiltonian and is instead transferred to the measurement stage through the ``POVM + postselection'' scheme. The operational cost of the CNOT-based measurement is the price for achieving interactions absent in the Hamiltonian.

\section{Simulacrum of violation of Bell inequality with unentangled particles}

Having introduced the entanglement simulacrum, we now turn to the recent experiment by Wang \textit{et al.}~\cite{WangZeilinger}, which reported a Bell violation with unentangled photons. This example is particularly relevant to the present framework because the observed correlations do not arise from a standard Bell test performed on entangled two-photon states. Rather, they emerge from a postselected subensemble whose fourfold-coincidence rates are normalized in a nonstandard way. We show below that such correlations can be understood as the manifestation of an alternative metric to $L^2$.

More specifically, the experiment realizes a frustrated-interference mechanism in which global correlations generated by the parametric operations are hidden by path identity~\cite{Herzog}. The setup consists of four coherently pumped probabilistic two-photon sources, each producing photon pairs in polarization product states. Two pairs of modes are sent to Alice and two to Bob, and by aligning the optical paths, Alice's pairs become mutually indistinguishable, and similarly for Bob. Phases $\varphi_A$ and $\varphi_B$ are applied to one on Alice's side and one on Bob's side, respectively. This removes any accessible which-source distinction—or, more precisely, prevents such information from existing in the first place—so that the indistinguishable pair-generation alternatives interfere coherently with one another, producing Bell correlations that the experiment attributes to path identity rather than to entanglement.

For fixed phase settings $(\varphi_A,\varphi_B)$, the detection events are classified according to the possible click and no-click patterns of the four detectors, and the analysis retains only the subensemble associated with simultaneous fourfold coincidences. The postselection discards all subensembles except the one in which all four detectors click simultaneously. Repeating the same postselected measurement for the shifted settings $ (\varphi_A+\pi,\varphi_B), (\varphi_A,\varphi_B+\pi), (\varphi_A+\pi,\varphi_B+\pi)$, and normalizing the four accepted fourfold-coincidence rates, produces the Bell correlation
\begin{equation}
E(\varphi_A,\varphi_B)=\cos(\varphi_A+\varphi_B), \label{BellCorr}
\end{equation}
which, for suitable choices of the phases, leads to a violation of a Bell inequality.

To explain the violation of Bell inequality without entangled particles, we adopt the same metric operator defined in Sec. V for the entanglement simulacrum: $\Theta=\eta^\dagger \eta$ with the Dyson map $\eta$ given by Eq.~(\ref{DMEnt}). Following the steps in Sec. V, we consider two spin-$1/2$ particles prepared in the product state $\ket{\psi}$ of Eq.~(\ref{EntIniSta}), to obtain the Bell-like state
\begin{equation}
\ket{\phi}=\eta\ket{\psi}=\cos\left(\frac{\varphi_{A}+\varphi_{B}}{2}\right)\frac{\left|\uparrow_{1}\uparrow_{2}\right\rangle +e^{i\Phi}\left|\downarrow_{1}\downarrow_{2}\right\rangle }{\sqrt{2}}+\sin\left(\frac{\varphi_{A}+\varphi_{B}}{2}\right)\frac{\left|\uparrow_{1}\downarrow_{2}\right\rangle +ie^{i\Phi}\left|\downarrow_{1}\uparrow_{2}\right\rangle }{\sqrt{2}},
\end{equation}
where $\Phi$ is defined by $\cos\Phi=-\cot^{2}\left(\frac{\varphi_{A}+\varphi_{B}}{2}\right)$. In order to establish the correspondence with the experiment, we choose the coefficients of the initial product state to reproduce the phase-dependent fourfold-coincidence probabilities. For simplicity, we evaluate the state at $t=0$, obtaining
\begin{subequations}
\begin{align}
a_{1}&=\sqrt{\frac{1-ie^{-i\Phi}}{1+ie^{-i\Phi}}}\left[\frac{\alpha}{2\beta}e^{-2\epsilon}\cos^{2}\left(\frac{\varphi_{A}+\varphi_{B}}{2}\right)\right]^{1/4}, \\
a_{2}&=\sqrt{\frac{1+ie^{-i\Phi}}{1-ie^{-i\Phi}}}\left[\frac{\beta}{2\alpha}e^{-2\epsilon}\cos^{2}\left(\frac{\varphi_{A}+\varphi_{B}}{2}\right)\right]^{1/4},\\
b_{1}&=e^{i\Phi/2}\left[\frac{\beta}{2\alpha}e^{+2\epsilon}\cos^{2}\left(\frac{\varphi_{A}+\varphi_{B}}{2}\right)\right]^{1/4}, \\
b_{2}&=e^{i\Phi/2}\left[\frac{\alpha}{2\beta}e^{+2\epsilon}\cos^{2}\left(\frac{\varphi_{A}+\varphi_{B}}{2}\right)\right]^{1/4}.
\end{align}
\end{subequations}
With this choice, the corresponding detection probabilities are
\begin{subequations}
\begin{align}
p_{\uparrow\uparrow}
=
p_{\downarrow\downarrow}
&=
\left|\langle\uparrow_{1}\uparrow_{2}|\phi\rangle\right|^{2}
=
\frac14
\left[
1+\cos(\varphi_A+\varphi_B)
\right],
\\
p_{\uparrow\downarrow}
=
p_{\downarrow\uparrow}
&=
\left|\langle\uparrow_{1}\downarrow_{2}|\phi\rangle\right|^{2}
=
\frac14
\left[
1-\cos(\varphi_A+\varphi_B)
\right].
\end{align}
\label{ProbBV}
\end{subequations}
Consequently, the correlation function becomes $ E(\varphi_A,\varphi_B) = p_{\uparrow\uparrow} + p_{\downarrow\downarrow} - p_{\uparrow\downarrow} - p_{\downarrow\uparrow}
= \cos(\varphi_A+\varphi_B)$, thereby reproducing Eq.~(\ref{BellCorr}) and the Bell-like statistics reported in Ref.~\cite{WangZeilinger}. The same probabilities derived above in the Hilbert space endowed with the metric $\Theta$ can be obtained through a postselected measurement associated with the biorthonormal basis $\{|\chi_n\rangle,|\psi_n\rangle\}$, with $n\in\{\uparrow\uparrow,\uparrow\downarrow,\downarrow\uparrow,\downarrow\downarrow\}$. The detection process is described by the POVM operators
$ E_n=q|\chi_n\rangle\langle\chi_n|$, together with the failure outcome
$ E_{\times}=\mathbbm{1}-q\Theta$, where $|\chi_n\rangle=\eta^\dagger|n\rangle$. After discarding the failure events associated with $E_\times$, the accepted probabilities, in accordance with Eq. (\ref{ProbBV}), are
\begin{equation}
p_n
=
\frac{
\langle\psi|E_n|\psi\rangle
}
{
\sum_m\langle\psi|E_m|\psi\rangle
}
=
|\braket{n|\psi}_\Theta|^2,
\end{equation}
where we have assumed the normalization condition $\braket{\psi|\psi}_\Theta=1$.

The application of our “POVM + postselection” measurement scheme to the Wang \textit{et al.} experiment requires the implementation of the accepted-event POVM operator $E_{\uparrow\uparrow}(\varphi_A,\varphi_B)$. The four coincidence rates entering the Wang \textit{et al.} protocol are then obtained as repeated realizations of this same accepted event, performed in distinct measurement runs with appropriately shifted phase settings. Defining the accepted coincidence rate for the setting $(\varphi_A,\varphi_B)$ as $\langle\psi|E_{\uparrow\uparrow}(\varphi_A,\varphi_B)|\psi\rangle$, the remaining rates are obtained from the POVM operators 
\begin{subequations}
\begin{align}
E_{\downarrow\uparrow}(\varphi_{A},\varphi_{B})
&=E_{\uparrow\uparrow}(\varphi_{A}+\pi,\varphi_{B}),
\\
E_{\uparrow\downarrow}(\varphi_{A},\varphi_{B})
&=E_{\uparrow\uparrow}(\varphi_{A},\varphi_{B}+\pi),
\\
E_{\downarrow\downarrow}(\varphi_{A},\varphi_{B})
&=E_{\uparrow\uparrow}(\varphi_{A}+\pi,\varphi_{B}+\pi).
\end{align}
\end{subequations}
As in the entanglement simulacrum discussed in Sec.~V, the measurements of the POVM operators listed above require a global operation, such as the CNOT gate used to define the biorthogonal basis associated with the metric. From this perspective, whereas the "POVM  + postselection" scheme in the entanglement simulacrum requires the implementation of a CNOT gate, which ultimately entangles the initially separable particles, in Ref.~\cite{WW} essentially the opposite occurs: the entanglement genuinely generated by the parametric down-conversion processes is rendered operationally irrelevant by the indistinguishability of the paths followed by the photons.

 Although the authors of Ref.~\cite{WW} correctly identified the combination of postselection and probability renormalization as the mechanism underlying the Bell-inequality violation with unentangled states, here we interpret this combination as the operational core of the Bell-violation simulacrum: a separable state in the Hilbert space endowed with the metric $\Theta$ reproduces the same postselected statistics that, in the usual $L^2$ metric, would be attributed to the Bell-like state $|\phi\rangle$. Thus, while the explanation given in Ref.~\cite{WW} correctly captures the mechanism responsible for the observed violation, it does not explicitly recognize the combined action of postselection and renormalization as defining an effective metric different from $L^2$. The realization that postselection and renormalization introduce new metrics places quantum simulacra within a broader theoretical framework, enabling new perspectives and opening the way to further applications as remarkable as the Bell-inequality violation with unentangled particles.

\section{Teleportation simulacrum}

We now consider a three-particle system to illustrate how a simulacrum of entanglement can be used to implement a quantum teleportation protocol \cite{QT}. We start from the bare Hermitian Hamiltonian $H_{123}=\omega_{1}\sigma_{1}^{z}+\omega_{2}\sigma_{2}^{z}+\omega_{3}\sigma_{3}^{z}$
together with the metric operator $\Theta=\eta^{\dagger}\eta\otimes I_{3}$, where the Dyson map $\eta$, the same as in Section V, acts only on qubits 1 and 2. The Hermitian counterpart of $H$ becomes $
h_{123}=h+\omega_{3}\sigma_{3}^{z}$, with $h$ describing an effective coupling between qubits 1 and 2, as given by Eq. (\ref{h-ent}). For simplicity, we focus on the maximum interaction $\zeta=1$.

We consider the initial separable state $
|\psi_{123}\rangle=\sqrt{\beta/\alpha}\;|\downarrow_{1}\uparrow_{2}\rangle\otimes|\psi_{3}\rangle$, normalized with respect to $\Theta$, where $|\psi_{3}\rangle=a|\uparrow_{3}\rangle+b|\downarrow_{3}\rangle$ is an arbitrary state of qubit 3. Mapping to the Hermitian representation via $|\phi_{123}\rangle=\eta\ket{\psi_{123}}$, it is convenient to introduce the Bell basis for qubits 2 and 3, $|\psi_{\pm}^{23}\rangle=\left(|\uparrow_{2}\downarrow_{3}\rangle\pm|\downarrow_{2}\uparrow_{3}\rangle\right)/\sqrt{2}$ and $|\phi_{\pm}^{23}\rangle=\left(|\uparrow_{2}\uparrow_{3}\rangle\pm|\downarrow_{2}\downarrow_{3}\rangle/\right)\sqrt{2}$. In this basis, the state $\ket{\phi_{123}}$ can be written as
\begin{equation}
|\phi_{123}\rangle=\sum_{B\in\{\psi_{\pm},\phi_{\pm}\}}
\left(a_{B}\ket{\uparrow_{1}}+b_{B}|\downarrow_{1}\rangle\right)\otimes|B^{23}\rangle /2,
\end{equation}
with coefficients $
(a_{\psi_{\pm}},b_{\psi_{\pm}})=(\pm a,\, b)$ and $(a_{\phi_{\pm}},b_{\phi_{\pm}})=(\mp b,\, \pm a)$. This is the standard teleportation decomposition in the Hermitian representation, in which Alice performs a Bell's measurement on qubits 2 and 3, while Bob holds qubit 1. Given the measurement outcome, qubit 1 collapses into a rotated version of the unknown state.

The standard Bell measure is related to the Hermitian observable $o_{23}$, whose eigenstates are the Bell states, given by $
o_{23}=\sigma_{2}^{x}\otimes\sigma_{3}^{x}+\sigma_{2}^{y}\otimes\sigma_{3}^{y}+\sigma_{2}^{z}\otimes\sigma_{3}^{z}$. In the pseudo-Hermitian representation, with $H$ and $|\psi_{123}\rangle$ under the metric $\Theta$, the corresponding observable is $O_{123} = \eta^{-1}\left(I_{1}\otimes o_{23}\right)\eta$, which describes how the Bell's measurement can be performed taking into account the $\Theta$-pseudo-Hermiticity. Expanding $|\psi_{123}\rangle$ in the biorthogonal basis defined by right and left eigenvector of $O_{123}$, given by $
|\psi_{s_{1},B}\rangle=\eta^{-1}\left(|s_{1}\rangle\otimes|B^{23}\rangle\right)$ and $
|\chi_{s_{1},B}\rangle=\eta^{\dagger}\left(|s_{1}\rangle\otimes|B^{23}\rangle\right)$, respectively, with $s_{1}\in\{\uparrow,\downarrow\}$ and $B\in\{\psi_{\pm},\phi_{\pm}\}$, we obtain
\begin{equation}
|\psi_{123}\rangle
=\sum_{B\in\{\psi_{\pm},\phi_{\pm}\}}
\left(a_{B}|\psi_{\uparrow,B}\rangle+b_{B}|\psi_{\downarrow,B}\rangle\right)/2.
\end{equation}
Following the POVM + postselection scheme described in Section III, we define the POVM operators $E_{s_{1},B}=q\ket{\chi_{s_{1},B}}\bra{\chi_{s_{1},B}}$ and $
E_{\times}=\mathbbm{1}-\sum_{s_{1},B}E_{s_{1},B}$, which enables the measurement of the observable \(O_{123}\) in the metric $\Theta$ via postselection, by discarding the inconclusive outcome associated with \(E_\times\). However, implementing the POVM operators requires the discrimination of the nonseparated states \(|\chi_{s_{1},B}\rangle\), which can be achieved by a two-control operation, the controls being particles \(2\) and \(3\), flipping the target \(1\) whenever the controls have even parity (both up or both down), i.e., a parity-controlled NOT gate, described by
\begin{equation}
\begin{aligned}
\mathrm{PCNOT}
&=\sigma_{1}^{x}\otimes\left(
|\downarrow_{2}\downarrow_{3}\rangle\langle\downarrow_{2}\downarrow_{3}|
+|\uparrow_{2}\uparrow_{3}\rangle\langle\uparrow_{2}\uparrow_{3}|
\right) \\
&\quad+I_{1}\otimes\left(
|\uparrow_{2}\downarrow_{3}\rangle\langle\uparrow_{2}\downarrow_{3}|
+|\downarrow_{2}\uparrow_{3}\rangle\langle\downarrow_{2}\uparrow_{3}|
\right).
\end{aligned}
\end{equation}

Upon successful postselection, Alice performs a standard Bell-state measurement on spins $2$ and $3$ using a CNOT gate. She then sends the measurement outcome to Bob, who applies the rotation  $R=\exp\left(-i\theta\hat{n}\cdot\overrightarrow{\sigma}/2\right)$ to spin 1, thereby recovering the state $|\psi_{1}\rangle=a|\uparrow_{1}\rangle+b|\downarrow_{1}\rangle$. The angles $\theta$ and $\phi$ are determined by equations
\begin{subequations}
\begin{align}
c_{2}\cos\theta+c_{1}e^{-i\phi}\sin\theta=b_B \mathcal{N}, \\
c_{2}e^{+i\phi}\sin\theta-c_{1}\cos\theta=a_B\mathcal{N},
\end{align}
\end{subequations}
where $c_1$ and $c_2$ denote the amplitudes of qubit $1$ associated with Alice’s Bell-state measurement outcome, resulting from the operation $\text{CNOT}\left(a_{B}\left|\Psi_{\uparrow,B}\right\rangle +b_{B}\left|\Psi_{\downarrow,B}\right\rangle \right)=\mathcal{N}\left(c_{1}\left|\downarrow_{1}\right\rangle +c_{2}\left|\uparrow_{1}\right\rangle \right)\otimes\ket{\tilde{B}^{23}}$, with $\mathcal{N}$ as the normalization constant and $\hat{n}=\left(\sin\theta\cos\phi,\sin\theta\sin\phi,\cos\theta\right)$. Therefore, we have described a protocol for a simulacrum of quantum teleportation induced by the metric $\Theta$: conditioned on the success of the ``POVM + post-selection" scheme, the unknown state encoded in the amplitudes $a$ and $b$ is transferred from qubit (3) to qubit (1), up to a rotation.

\section{Superradiant phase transition from metric}

To construct a simulacrum of the Tavis--Cummings interaction, we consider an ensemble of $N$ identical two-level emitters and a single bosonic mode, with the bare Hermitian Hamiltonian
\begin{equation}
H=\omega_{0}S_{z}+\omega\left(a^{\dagger}a+1/2\right), \label{H-SR}
\end{equation}
where $S_{z}=\sum_{n=1}^{N}\sigma_{z}^{(n)}/2$ and $S_{+}=\sum_{n=1}^{N}\sigma_{+}^{(n)}=S_{-}^{\dagger}$ are the collective spin operator. Since the total excitation number $\mathbf{N}=a^{\dagger}a+S_{z}+N/2$ is conserved, the Hilbert space decomposes into invariant subspaces labeled by $\mathbf{N}$. Within each subspace, it is convenient to introduce the collective operators $J_{+}=aS_{+}/\sqrt{\mathbf{N}}$, $J_{-}=a^{\dagger}S_{-}/\sqrt{\mathbf{N}}$ and $J_{z}=S_{z}$, which close an $\mathfrak{su}(2)$ algebra. We then define the Dyson map $\eta=\exp\left[\epsilon S_{z}+\chi\left(\alpha aS_{+}-\beta a^{\dagger}S_{-}\right)\right]$, where $\epsilon$ and $\chi$ are functions of the real parameters $\alpha$, $\beta$, and $z$, given by
\begin{subequations}
\begin{align}
\epsilon(z) & =-\frac{1}{\sqrt{1-z^{2}}}\tanh^{-1}\left[\frac{\alpha-\beta}{\alpha+\beta}\sqrt{1-z^{2}}\right], \\
\chi(z) &= \frac{\epsilon(z)}{\sqrt{\mathbf{N}\alpha\beta}}.
\end{align}
\end{subequations}
For each choice of $z$ in the interval $-1<z<1$ and parameters $\alpha$, $\beta$ satisfying $\alpha\beta>0$, the Hermitian counterpart of $H$ takes the Tavis--Cummings model form,
\begin{equation}
h=\left[\omega_{0}-2\Delta\zeta^{2}/\left(1+\zeta^{2}\right)\right]S_{z}+\omega\left(a^{\dagger}a+1/2\right)+g\left(aS_{+}+a^{\dagger}S_{-}\right), \label{h-SR}
\end{equation}
where $\Delta=\omega_{0}-\omega$ is the detuning and
$\zeta=(\alpha-\beta)z/2\sqrt{\alpha\beta}$ is a metric parameter that determines the coupling $g=2\Delta\zeta/\sqrt{\mathbf{N}}\left(1+\zeta^{2}\right)$. As in the previous examples, here the light-matter interaction is absent from the Hamiltonian $H$,  emerging in $h$ through a suitable choice of metric.  

In the Tavis--Cummings model with the standard metric, an equilibrium superradiant phase transition requires the effective coupling $\lambda=\sqrt{N}g$ to exceed the critical value $\lambda_c=\sqrt{\omega_{0}\omega}$ \cite{Wang}. This condition places $\lambda$ on the order of the atomic and cavity frequencies, corresponding to a very strong-coupling regime \cite{Dimer}. However, the Tavis--Cummings Hamiltonian is obtained from the Dicke model \cite{Duncan} by applying the rotating-wave approximation (RWA), whose validity requires $\lambda$ to be much smaller than both $\omega$ and $\omega_0$. Therefore, the critical coupling cannot be reached without leaving the RWA regime, making the equilibrium superradiant phase transition inaccessible within the standard $L^2$ metric under this approximation \cite{Kirton}. The simulacrum of the Tavis--Cummings interaction circumvents this limitation because the effective Hamiltonian in Eq.~(\ref{h-SR}) is not obtained by applying the RWA to the Dicke model. More precisely, there exists a family of metrics $\Theta=\eta^\dagger\eta$, parametrized by $\zeta$ in the interval $[\zeta_{-},\zeta_{+}]$, with
\begin{equation}
\zeta_{\pm}=\Delta\sqrt{N}/\omega_{0}\pm\sqrt{\left(\Delta\sqrt{N}/\omega_{0}\right)^{2}-1},
\end{equation}
that yields an effective coupling $\lambda$ above the critical value. These metric choices therefore make the transition from the normal phase to the superradiant phase feasible, provided that the corresponding pseudo-Hermitian observables can be measured. An example is the photon number $\hat{n}=\eta^{-1} \left(a^\dagger a\otimes \mathbbm{1}\right) \eta$, which may be accessed through our ``POVM + postselection" scheme or through alternative measurement protocols. This metric-induced superradiant phase transition constitutes a genuine simulacrum: the phenomenon arises from a suitable choice of metric and has no counterpart under the standard $L^2$ metric.

\section{Other examples of quantum simulacra}

In addition to the examples discussed above, several results already reported in the literature \cite{Ponte, Ponte2, Cius, Cius2} also admit a natural interpretation as quantum simulacra. A first example is provided by time-dependent $\mathcal{PT}$-symmetric bosonic models for a mode of the electromagnetic field endowed with time-dependent metrics. Starting from all-annihilation Hamiltonians, whose standard $L^2$ description contains no creation terms, authors of Ref. \cite{Ponte} obtain an effective parametric-amplification mechanism that leads to an infinite degree of squeezing at a finite time. This behavior has no counterpart in Hermitian models under the standard metric: squeezing may increase over time, but it reaches an infinite degree only asymptotically.

Another example is the enhancement of creation of photons in cavities from the vacuum in the dynamical Casimir effect \cite{Cius}. In its standard formulation, this effect requires large and rapid modulations of a cavity length or frequency, making its observation experimentally demanding. In the pseudo-Hermitian formulation of Ref. \cite{Cius}, an appropriate choice of metric amplifies the photon-production rate, offering a possible route to observing Casimir photons without relying on such rapid modulations. Related examples include the generation of entanglement between otherwise uncoupled modes driven by a time-dependent complex frequency \cite{Cius2}. These examples are as genuine quantum simulacra because the corresponding amplification or correlation channels are absent in the standard-metric description.

The cost of this construction is the measurement of metric-redefined observables, which can be implemented through suitable POVMs combined with postselection, as proposed here. In this sense, the interactions or amplification channels absent from the Hamiltonian are transferred to the measurement stage. Although some of these examples remain proofs of principle, they show that nontrivial couplings can be engineered through the choice of metric even when the underlying Hamiltonian contains no corresponding interaction term.

\section{Conclusions}

Building on the metric structure of pseudo-Hermitian quantum mechanics, we have shown how the same framework can be extended to systems described by Hermitian Hamiltonians, for which the metric commutes with the Hamiltonian and is therefore associated with the invariant and symmetry structure of the problem. Within this extended framework, we have introduced the concept of quantum simulacra as phenomena that arise when a suitable choice of metric gives access to processes not encoded in the Hamiltonian itself. By redefining the set of observables and, consequently, the measurement context, the metric becomes an additional component in the operational description of microscopic reality. In this sense, the Hamiltonian alone does not exhaust the physical content of the theory; the metric specifies how the system is interrogated and which effective processes become observable.

This Hamiltonian--metric connection is illustrated by the examples developed here. A free field mode acquires squeezing, uncoupled qubits display entanglement and support teleportation, separable states reproduce Bell-violating statistics, and a system without an explicit light--matter interaction can undergo a superradiant phase transition. Although they arise in distinct physical settings, these examples share the same underlying principle: a suitable metric can make accessible, within an operational description based on "POVM + postselection", phenomena that would be unavailable in the standard $L^2$ metric.

The price of this metric-induced plasticity is the need to implement measurements of the redefined observables. To address this point, we proposed the “POVM + postselection” scheme that implements the measurements associated with the required biorthogonal basis within the accepted subensemble, thereby reconstructing the statistics associated with the selected metric. Quantum simulacra thus shift part of the usual burden of engineering dynamics from the Hamiltonian to the measurement stage. This perspective suggests a route for realizing effective quantum processes not by adding interactions to the system, but by designing the metric structure through which the system is measured. We expect that further analyses of quantum simulacra may reveal potential advantages in quantum communication and computation, two fields that have been extensively investigated over the past decades.

\begin{flushleft}
\textbf{{\Large Acknowledgements}}
\end{flushleft}

The authors acknowledge financial support from CAPES, CNPq, and FAPESP, Brazilian agencies.

\end{document}